\def\pd#1#2{ { \partial #1 \over \partial {#2} } }
\def\pdd#1#2{ { \partial^2 #1 \over \partial {#2}^2 } }

\def\half{\frac{1}{2}}

\newcommand{\labeq}[2]{ \begin{equation} \llabel{#1} #2 \end{equation}}

\newcommand{\llabel}[1]{\label{#1}}                 

\documentclass[11pt]{article}      
\usepackage{amstext,amsmath}
\usepackage{graphicx}

\title{\bf A scalar hyperbolic equation \\ with GR-type non-linearity}  

\author{A.M Khokhlov\footnote{Laboratory for Computational Physics, Code 6404, Naval Research Laboratory, Washington, DC 20375}
              and  
        I.D. Novikov\footnote{Theoretical Astrophysics Center, Juliane Maries vej 30, DK-2100 Copenhagen, Denmark}
                    \footnote{Copenhagen University Observatory, Juliane Maries vej 30, DK-2100 Copenhagen, Denmark}
                    \footnote{Astro Space center of the P.N. Lebedev Physical Institute, Profsoyouznaja 84/32, Moscow 118710, Russia}
                    \footnote{NORDITA, Blegdamsvej 17, DK-2100 Copenhagen, Denmark} 
       }

\begin{document}

\maketitle                   

\large

\begin{abstract}
We study a scalar hyperbolic
partial differential equation with non-linear terms similar to
those of the equations of general relativity. 
The equation has a number of non-trivial analytical solutions
whose existence rely on a delicate balance between linear and non-linear
 terms. 
We formulate two classes of second-order accurate central-difference
 schemes, CFLN and MOL,  for numerical integration of this equation.
Solutions produced by the schemes converge to exact solutions
at any fixed time $t$ when numerical resolution is increased.
However, in certain cases integration becomes 
asymptotically unstable when $t$ is increased and
 resolution is kept fixed. 
This  behavior is caused by subtle changes in the balance between 
linear and non-linear terms when the equation is discretized in space.
Changes in the balance occur without violating a second-order accuracy of the
discretization.
We thus demonstrate that second-order accuracy and convergence at finite $t$ does not
guarantee a  correct asymptotic behavior and long-term numerical stability.
 Accuracy and long-term stability of integration are greatly improved
 by an exponential transformation  of the unknown variable.   
\end{abstract}

\clearpage

\numberwithin{equation}{section}

\section{Introduction} 
It is well known that numerical integration of Einstein's equations
 in a 3+1 form often leads to
instabilities and terminates prematurely.
Some of the instabilities have been related to the 
 presence of 
constraints, and some to the existence of 
 gauge degrees of freedom in Einstein's  equations (\cite{SHINKAI-02},\cite{KHOKHLOV-02}, and references therein).
These instabilities are intrinsic to the equations of
 general relativity (GR) themselves.
Numerical instabilities may also arise due to a bad choice of a 
finite-difference scheme.

In this paper we introduce a scalar hyperbolic
partial differential equation with non-linear terms similar to
those of much more complicated equations of GR (Section 2).
The equation has 
a number of non-trivial analytical solutions (Sections 3 and 4). 
We experiment with several finite-difference schemes for
 numerical integration of this
equation (Section 5).

Solutions produced by the schemes converge to exact 
solutions at any fixed moment
of time $t$ when numerical resolution is increased.
However, 
in certain cases numerical integration becomes
unstable 
when $t$ is increased while the resolution 
is kept fixed (Section 6).  We trace this
 behavior to a special structure of the
non-linear terms and demonstrate that discretization of non-linear terms
 leads to a finite-difference system whose
asymptotic behavior may differ qualitatively from the corresponding 
behavior of a continuous system (Section 7).  
 The accuracy and stability of integration 
can be greatly improved
 by an exponential transformation  of the unknown 
variable (Section 7).  

\section{A model equation}

Consider a 
 scalar, quasi-linear, hyperbolic partial
differential equation of two independent variables, $t$ and $x$, 
\labeq{SCALAR}{a_{11} g_{tt} + 2 a_{12} g_{tx} + a_{22} g_{xx} + g^{-1} ( b_{11} g_t^2 + 2 b_{12} g_t g_x + b_{22} g_x^2 ) = 0,
}
where $g=g(t,x)$ is the unknown, 
and $a_{ij}$, $b_{ij}$ are constant coefficients, 
$a_{11} a_{22} - a_{21}^2 < 0$. 
The non-linear term in \eqref{SCALAR} has the form 
$g^{-1}\,\sum\text{(first derivatives squared)}$.

The reason for considering \eqref{SCALAR}
 becomes obvious if we recall that
the structure of a Ricci tensor $R_{ab}$ is
$R \sim \sum\partial\Gamma + \sum \Gamma \Gamma$, 
where Cristoffel symbols $\Gamma \sim
g^{-1}\partial g$, and $g$ is the metric. 
Thus, $R_{ab}$ is represented as a sum of the terms 
$R \sim \sum g^{-1} \partial^2 g + \sum
g^{-2} (\partial g)^2$. Equation \eqref{SCALAR} thus mimics
a type of non-linearity present in GR equations $R_{ab}=0$.
In particular, \eqref{SCALAR} may have a singularity related to $g$ 
becoming zero.

Equation \eqref{SCALAR} can be
reduced to its normal form,
\labeq{SIMPLE}{
g_{tt} = g_{xx} - g^{-1} ( \alpha g_t^2 + \beta g_x^2 + \gamma g_x g_t),
}
by a  linear transformation which preserves a
 quadratic form of the non-linearity.
We work below with a simpler equation \eqref{SIMPLE}.
This is a non-linear hyperbolic equation with a characteristic speed equal to 
$1$.

By introducing a new variable $K = g_t$, 
we rewrite \eqref{SIMPLE} as a system of two first-order in time 
partial differential equations
\labeq{SIMPLE0}{
\begin{split}
 & g_t = K, \\
 & K_t =  g_{xx} - g^{-1} ( \alpha K^2 + \beta g_x^2 + \gamma K g_x)\\
\end{split}
}
which resembles the evolutionary part of GR equations in 
a standard ADM 3+1 form (with zero shift and constant lapse).
By introducing yet another variable $D = g_x$
we further rewrite \eqref{SIMPLE} as a system of first-order PDE
\labeq{SIMPLE1}{
\begin{split}
 g_t &= K,\\
 K_t - D_x &=  R, \\
 D_t - K_x &= 0,\\
\end{split}
}
where
\labeq{R}{R \equiv - g^{-1} ( \alpha K^2 + \beta D^2 + \gamma D K).
}
This system has a complete set of real eigenvalues and eigenvectors.
%

\section{Exponential transformation}

A transformation
\labeq{SUBEXP}{
                    g = e^\phi,
}
where $\phi$ is a new unknown, will
play an important role in subsequent sections. 
After substitution of \eqref{SUBEXP} into \eqref{SIMPLE} 
we obtain a PDE for a new unknown $\phi$,
\labeq{SIMPLEX}{
\phi_{tt} = \phi_{xx} - (\alpha + 1) \phi_t^2 - (\beta - 1) \phi_x^2 - \gamma \phi_x \phi_t.
}
The transformation \eqref{SUBEXP} removes $g^{-1}$ multiplier in front of the non-linear term in \eqref{SIMPLE}, and
maps $0 < g  < \infty$ onto $-\infty < \phi  < \infty$ so that values of 
 $g \leq 0$ are excluded. This is consistent with  GR where three-dimensional 
metric of space-like hypersurfaces is positive-definite.

We can introduce new variables 
\labeq{LOGVARS}{\psi \equiv \phi_t, \quad \theta \equiv \phi_x,
}
and rewrite \eqref{SIMPLEX} in the equivalent two-equation form similar
 to \eqref{SIMPLE0}
as 
\labeq{SIMPLEX1}{
\begin{split}
 \phi_t &= \psi,\\
 \psi_t &= \phi_{xx} - (\alpha + 1) \psi^2 
    - (\beta - 1) \phi_x^2 
    - \gamma \psi \phi_x ,\\
\end{split}
}
or in a three-equation form similar to \eqref{SIMPLE1} as
\labeq{SIMPLEX2}{
\begin{split}
 \phi_t &= \psi,\\
 \psi_t - \theta_x &=  S, \\
 \theta_t - \psi_x &= 0,\\
\end{split}
}
where
\labeq{SIMPLEXS}{S = - (\alpha + 1) \psi^2 - (\beta - 1) \theta^2 - \gamma \psi \theta. }
System \eqref{SIMPLEX2}obviously has the same eigenvectors and 
eigenvalues as
\eqref{SIMPLE1}.

\section{Analytic solutions}

For a choice of parameters
\labeq{LIN}{\alpha=-1,~ \beta=1, ~\text{and}~ \gamma=0,
}
 \eqref{SIMPLEX} becomes a linear hyperbolic PDE 
\labeq{LIN1}{\phi_{tt}=\phi_{xx}}
whose general solution  is
\labeq{LIN1SOL}{ \phi = \phi_1 ( x+ t ) + \phi_2(x-t),}
where $\phi_{1,2}$ are arbitrary functions.
The original equation \eqref{SIMPLE} remains non-linear,
\labeq{LIN2}{ g_{tt} - g_{xx} =  g^{-1} ( g_t^2 - g_x^2 ).
}
Its general solution is
\labeq{LIN2SOL}{
g = g_1(x+t) \cdot g_2(x-t),
}
where $g_1$, $g_2$ are arbitrary functions.

For
$\alpha$, $\beta$, and $\gamma$ other than
 \eqref{LIN}, equation \eqref{SIMPLEX} is non-linear but
we can find particular solutions of this equation
 in a form of a wave running with a constant speed $a$,
\labeq{XAT}{\phi = \phi(\zeta), \quad \zeta = x+at. }
Substituting \eqref{XAT} into \eqref{SIMPLEX} we obtain an
 ordinary differential equation
\labeq{SIMPLEF}{ (a^2-1) \frac{d^2\phi}{d\zeta^2} 
       + b \left(\frac{d\phi}{d\zeta} \right)^2 =0,
}
where
\labeq{b1}{ b = (\alpha+1) a^2 + \beta-1 + \gamma a. 
 }
We need to solve this equation for $\phi$.

We consider 
separately the cases $b(a^2-1) \neq 0$, $b=0$, and
$a^2=1$.
If $b(a^2-1) \neq 0$, integration of
\eqref{SIMPLEF} gives 
\labeq{anot1b}{
\phi =
            p\, \ln (c_0 + x + a t ) + c_1 ,\quad {\rm where}\quad p= \frac{a^2-1}{b},
}
and $c_0$ and $c_1$ are arbitrary constants. The corresponding solution of
\eqref{SIMPLE} is
\labeq{ANOT1B}{
g = e^{c_1} \left( c_0 + x + a t \right)^p.
}
If $a^2 \neq 1$ and $b=0$, which will happen if we chose 
\labeq{}{a = \frac{-\gamma \pm \sqrt{\gamma^2 
+ 4 (\alpha+1)(1-\beta)}}{2(\alpha+1)} \neq \pm 1,
}
 then integration of \eqref{SIMPLEF} gives 
\labeq{anot1a}{\phi = c_0 ( x + a t ) + c_1,
}
and the corresponding solutions of \eqref{SIMPLE} are
\labeq{ANOT1A}{ g = e^{c_0 ( x + a t ) + c_1 }.
}
Finally, if $a^2=1$, integration of \eqref{SIMPLEF} gives 
\labeq{a1}{
\phi = \left\{ \begin{array} {ll}    
             ~\phi_1(x+at) & \textrm{if $b=\alpha+\beta+ a\gamma=0$} \\
            const & \textrm{if $b=\alpha+\beta+ a\gamma\neq 0$} \\
            \end{array}
    \right. ,
}
where $\phi_1$ is arbitrary.
The speed $a$ of solutions \eqref{ANOT1B}, \eqref{ANOT1A} differs 
from the characteristic speed of \eqref{SIMPLE}.
These running wave solutions exist as a result of a delicate 
balance of linear and non-linear terms in \eqref{SIMPLE}.

\section{Numerical schemes}

Let us now consider how  \eqref{SIMPLE}
may be solved numerically.
Without non-linear terms, \eqref{SIMPLE} is a 
scalar wave equation $g_{tt} = g_{xx}$. 
A classical, central-difference,
 second-order accurate scheme for this equation  was introduced in 1928 by
 Courant, Friedrichs and Levy (see \cite{CFL1928} and  chapter 10 in  \cite{RM}),
\labeq{CFL}{
\frac{g^{n+1}_i - 2 g^n_i + g^{n-1}_i}{\Delta t^2} = 
\frac{g^n_{i+1} - 2 g^n_i + g^n_{i-1}}{\Delta x^2},
}
where $g_i^n$ are determined at mesh points 
$x_i = i\Delta x$ and $t^n = n \Delta t$.
The scheme is stable for Courant numbers
$cfl=\frac{\Delta t}{\Delta_x} \leq 1$.

We now proceed to expand \eqref{CFL} to the non-linear equation. 
We cast  \eqref{CFL} into a numerically equivalent 
 two-equation form,
\labeq{CFL1}{
\begin{split}
& \frac{K^{n+\half}_i - K^{n-\half}_i}{\Delta t} =
   \frac{g^n_{i+1} - 2 g^n_i + g^n_{i-1}}{\Delta x^2}, \\
& \frac{g^{n+1}_i - g^n_i}{\Delta t} = K^{n+\half}_i, \\
\end{split}
}
and extend it to \eqref{SIMPLE0} by adding
a discretized 
non-linear term 
to the right-hand side of the first
equation in \eqref{CFL1}. In order to maintain 
an overall second-order accuracy of the scheme, we must add the non-linear
 term  evaluated
with second-order accuracy at grid points $(x_i,t^n)$. This
 requires second-order accurate values of $g$ and $K$ at these points.
Values of $g_i^n$ are already defined there
but $K^n_i$ are not.
Taking $K^n_i = \half(K^{n+\half}_i + K_i^{n-\half})$ will give us the
desired accuracy but will also
render the scheme implicit. 
To escape this difficulty, we use  a 
predictor-corrector approach.

We write a discretized
 non-linear term \eqref{R} as
\labeq{R1}{
{\cal R}(g_i,K_i,D_i) = -g_i^{-1} \left( \alpha K_i^2 
         + \beta D_i^2
         + \gamma K_i D_i
\right), 
}
where $D_i = \frac{1}{2\Delta x}(g_{i+1}-g_{i-1})$,
without explicitly specifying a moment of time at which $g_i$ and $K_i$, 
must be taken. Using this notation, we  
write a second-order accurate 
explicit predictor-corrector scheme for
\eqref{SIMPLE0} as
\labeq{CFLN1}{ {\rm CFLN1:}\quad\quad \left\{
\begin{split}
\quad & \bar K_i = K^{n-\half}_i +
  \Delta t \left(\frac{g^n_{i+1} - 2 g^n_i + g^n_{i-1}}{\Delta x^2}
   + {\cal R}(g^n_i,K^{n-\half}_i,D_i^n)\right) \quad(\rm predictor), \\
\quad &  K_i^{n+\half} = \bar K_i + 
               \frac{ \Delta t}{2} \left( {\cal R}(g_i^n,\bar K_i,D^n_i)
                       - {\cal R}(g^n_i,K^{n-\half}_i,D^n_i)\right)
        \quad (\rm corrector) , \\
\quad & g^{n+1}_i = g^n_i + \Delta t \, K^{n+\half}_i,\\
\end{split}
\right.
}
We will refer to \eqref{CFLN1} as to a CFLN1 scheme.
Initial conditions for the scheme must be provided at $t=t^0$
 for $g_i$ and at $t = t^0-\half \Delta t$ for $K_i$.
Boundary conditions are required only for $g$ and must be provided at
$t^n$. We show elsewhere that the scheme is stable for $cfl \leq 1$
according to a standard VonNeuman stability analysis \cite{JACOB}. 
Thus, it may be  expected to converge 
 at any $t$ to exact solutions when the resolution is increased, 
$\Delta x \rightarrow 0$.
Numerical experiments presented in the next section confirm this assertion.

Note, that CFL1 can be easily cast into a  three-equation form, 
a numerical counterpart of  \eqref{SIMPLE1} 
\labeq{CFLN2}{ {\rm CFLN2:}\quad\quad \left\{
\begin{split}
\quad & \frac{\bar K_i - K^{n-\half}_i}{\Delta t} =
   \frac{D^n_{i+\half} - D^n_{i-\half}}{\Delta x}
+ {\cal R}(g^n_i,K^{n-\half}_i,D^n_i) \quad(\rm predictor), \\
\quad &  K_i^{n+\half} = \bar K_i + 
               \half \Delta t \left( {\cal R}(g_i^n,\bar K_i, D_i^n)
                       - {\cal R}(g^n_i,K^{n-\half}_i,D^n_i)\right)
        \quad (\rm corrector) , \\
\quad & \frac{D_{i+\half}^{n+1} - D_{i+\half}^n}{\Delta t} 
  = \frac{K^{n+\half}_{i+1} - K^{n+\half}_i}{\Delta x}, \\
\quad &  \frac{g^{n+1}_i - g^n_i}{\Delta t} = K^{n+\half}_i,\\
\end{split}
\right.
}
where we introduced new quantities
\labeq{D}{
D_{i+\half} = \frac{g_{i+1} - g_i}{\Delta x},
}
so that $D_i = \frac{1}{2}(D_{i+\half} + D_{i-\half})$.
We will refer to \eqref{CFLN2} as to
a CFLN2 scheme.
 Schemes CFLN1 and CFLN2 
are numerically equivalent and have identical  stability 
and convergence properties. 

A second group of numerical schemes considered in this paper is based on 
a method-of-lines approach.
We discretize \eqref{SIMPLE0} in space
using central differences, 
\labeq{MOL1}{ {\rm MOL1(n):} \quad\left\{
\begin{split}
\quad & \pd{g_i}{t} = K_i, \\
\quad & \pd{K_i}{t} = \frac{g_{i+1} + g_{i-1} - 2 g_i}{\Delta x^2} 
              + {\cal R}(g_i,K_i,D_i),
\end{split}
\right.
}
and integrate the
resulting system of ordinary differential equations in time 
using Runge-Kutta methods of orders $n=2,3,4$.
In what follows,
we refer to \eqref{MOL1} as to an MOL1$(n)$ scheme. The scheme is
second-order accurate and stable for sufficiently small Courant numbers.
Initial conditions for $g_i$ and $K_i$ must be provided
at $t=t^0$. Boundary conditions are required for $g_i$
at times  $t^n$.

Instead of \eqref{SIMPLE0}, we can 
discretize \eqref{SIMPLE1} in space,
\labeq{MOL2}{ {\rm MOL2:} \quad\left\{
\begin{split}
\quad & \pd{g_i}{t} = K_i, \\
\quad & \pd{D_{i+\half}}{t} = \frac{K_{i+1} - K_i}{\Delta x}, \\
\quad & \pd{K_i}{t} = \frac{D_{i+\half} - D_{i-\half}}{\Delta x} 
              + {\cal R}(g_i,K_i,D_i),
\end{split}
\right.
}
and then use Runge-Kutta methods of order $n$ 
to integrate \eqref{MOL2} in time. 
We will refer to \eqref{MOL2} as to a MOL2$(n)$ scheme.
Using \eqref{D}, it is  easy to verify 
that MOL1 and MOL2 schemes are in fact
equivalent.

Instead of a Runge-Kutta, one can use 
any other stable ODE integrator in MOL1 and MOL2, e.g., an iterative
Crank-Nicholson (ICN) scheme with appropriate number of iterations
\cite{TEUKOLSKY-00}. 
We do not present here results obtained
with the ICN as they are similar to those obtained with 
both CFLN1 and Runge-Kutta MOL1 schemes, and do not change any of the
conclusions of the paper. 
%

\section{Numerical convergence and asymptotic stability}

In this section, we present examples of numerical integration 
of \eqref{SIMPLE0}.
They illustrate convergence of the schemes at 
fixed time $t$ when the resolution is increased, $\Delta x\rightarrow 0$. 
 The examples also illustrate numerical difficulties which may arise when 
$\Delta x$ is kept fixed and integration time 
is increased, $t\rightarrow \infty$.

As a first example,  consider equation \eqref{SIMPLE} with
$\alpha = -\frac{1}{2}$, $\beta=\frac{5}{4}$, $\gamma = 0$.
We pick a wave speed $a=2$ and find from \eqref{b1}, \eqref{anot1b},
\eqref{ANOT1B} a particular growing solution 
\labeq{E1}{
{\rm E1}: \quad g = \left( x + 2 t \right)^{\frac{4}{3}}.
}
For $a=0.1$ we find a decaying solution
\labeq{E2}{
{\rm E2}: \quad g = \left( x + \frac{t}{10} \right)^{-3.88}.
}

We wish to obtain solutions E1 and E2 numerically
on interval $0.1  \leq x \leq 1.1$, and $t > 0$  using
 $N$ grid points with coordinates 
\labeq{}{
x_i = 0.1 + \Delta x \, (i-1/2)~,~~\Delta x = 1/N,
}
and boundary conditions
\labeq{BOUND1}{
g_{-\half}^n = g(0.1-\frac{\Delta x}{2}, n \Delta t), \quad
g_{N+\half}^n = g(1.1+\frac{\Delta x}{2}, n \Delta t), 
}
where $g$ is the corresponding exact solution (E1 or E2). 
Results of numerical integration 
are presented in Tables 1 and 2, and in Figure 1.

Table 1 shows convergence of a MOL1(4) scheme for a growing solution E1.
The maximum error norm $L_\infty$ indicates a second-order 
convergence. Relative error of integration decreases with time. 
It is possible to continue stable integration of E1
until the limit of large numbers is reached in a computer.
Results for MOL1(2), MOL1(3), and CFLN1 schemes are similar.

Table 2 illustrates convergence of a MOL1(4) scheme for a decaying solution 
E2. Similar to E1, convergence is second-order.
However, the relative error is much larger than in the E1 case
 and it grows with time.


\bigskip
\begin{table}  [!hbp]


\centerline{
\begin{tabular}{|r|r|r|}
\hline
N &  $L_{\infty}(t_1)$ & $L_{\infty}(t_2)$ \\
\hline
16  & 2.4E-04 & 7.9E-05 \\
32  & 4.7E-05 & 1.4E-05 \\
64  & 1.2E-05 & 5.9E-06 \\
128 & 3.0E-06 & 1.6E-06 \\
256 & 7.5E-07 & 4.2E-07 \\
\hline
\end{tabular}
}

\caption{Convergence of MOL1(4) numerical scheme for a growing solution E1.
Integration is carried out with a 
time step $\Delta t = \half \Delta x$ 
for $0.1 \leq x \leq 1.1$ and $t \ge 0$. The error norm $L_\infty =
\max_i\,\vert g_i/g_e-1\vert$, where $g_e$ is the exact solution 
\eqref{E1}, is given for two moments of time, $t_1=9.9$ and $t_2=24.75$.
}

\bigskip

\end{table}

Figure 1 compares numerical solutions in the middle
of the interval, $x=0.6$, to the exact
solution E2 at this point. For $t \leq 20$, the $N=2048$ numerical solution 
and the exact solution
cannot be distinguished on the plot.
The  $N=128$ numerical solution shows large deviations thich
 grow with time. For $N=64$, the deviations are so violent that the code
crashes well before reaching $t=20$. 
Results obtained using MOL1(2), MOL1(3),
and CFLN1 are similar. By increasing $N$
we can achieve a convergent solution at any fixed time $t$. 
However, if we keep the resolution fixed and increase $t$, we find that
numerical errors make a long-term stable integration impossible. 

\bigskip
\begin{table}  [!hbp]


\centerline{
\begin{tabular}{|r|r|r|}
\hline
N &  $L_{\infty}(t_1)$ & $L_{\infty}(t_2)$ \\
\hline
64  & 2.9E-01 & NaN \\
128 & 1.3E-01 & 2.2E-01 \\
256 & 3.0E-02 & 8.1E-02 \\
512 & 7.5E-03 & 1.9E-02 \\
1024 & 1.9E-03 & 4.8E-03 \\
2048 & 4.7E-04 & 1.2E-03 \\
\hline
\end{tabular}
}

\caption{Convergence of MOL1(4) numerical scheme for a decaying solution E2.
Integration is carried out with a 
time step $\Delta t = \half \Delta x$ 
for $0.1 \leq x \leq 1.1$ and $t \ge 0$. The error norm $L_\infty =
\max_i\,\vert g_i/g_e-1\vert$, where $g_e$ is the exact solution 
\eqref{E1}, is given for two moments of time, $t_1=9.9$ and $t_2=24.75$.
}

\bigskip

\end{table}

\bigskip
\begin{figure}
\centering
\vskip -1 cm
\includegraphics[totalheight=4in]{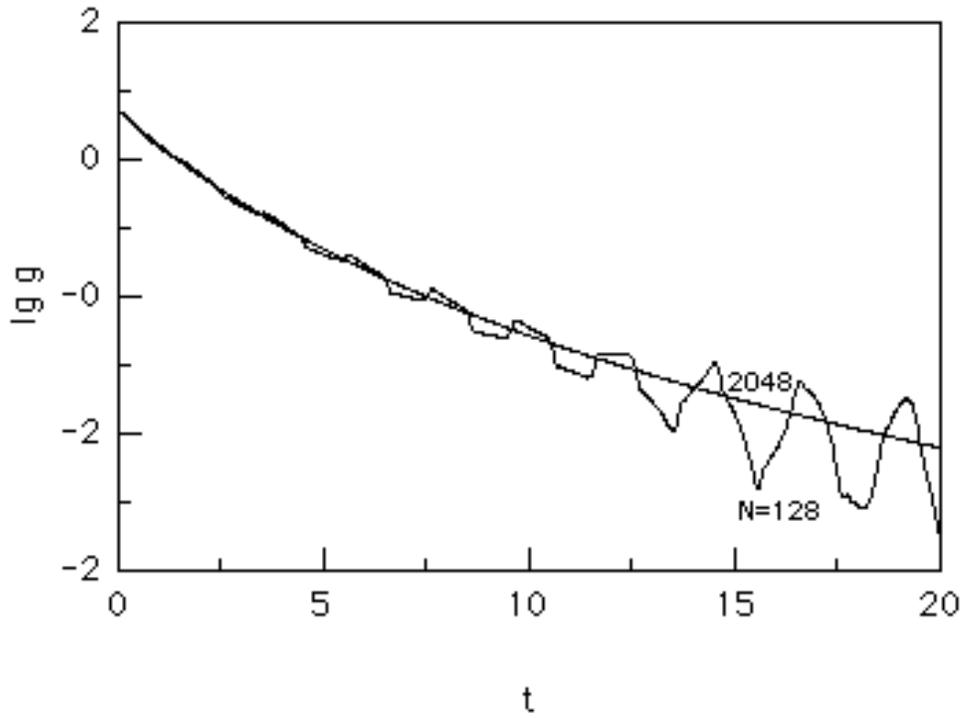}
\bigskip
\bigskip
\bigskip
\bigskip
\caption{A decaying solution \eqref{E2} at $x=0.6$ as a function of time.
Obtained using MOL1(4) scheme
 and $cfl=1/2$. Solid lines - numerical solutions for $N=128$ and  $N=2048$.
The exact and $N=2048$ numerical solutions cannot be distinguished on this plot.}
\end{figure}

To understand this behavior of numerical schemes, 
it is instructive to consider a special case of
exponential solutions \eqref{ANOT1A} which 
can be investigated analytically.
As an example, we take $\alpha =1$, $\beta=\gamma=0$ in \eqref{SIMPLE0}.
This choice of parameters gives 
a pair of exponential solutions \eqref{ANOT1A} 
with the speed $a=\pm \frac{1}{\sqrt{2}}$,
\labeq{E3}{
g_\pm = \exp\left( x \pm \frac{t}{\sqrt2}\right).
}
We now attempt to obtain $g_\pm$  numerically on 
the interval $0\leq x\leq 1$ discretized using $N$ grid points, 
\labeq{}{
x_i = \frac{1}{N} \left( i - \half \right) , \quad i = 1,...,N,
}
and applying Robin boundary conditions $\pd{\ln g}{x}=1$ for $i=0$ and $i=N+1$,
\labeq{ROBIN}{
   g_0 = \exp ( \ln g_2 - 2 \Delta x \ln g_1), 
   \quad g_{N+1} = \exp ( \ln g_{N-1} + 2 \Delta x \ln g_N).
}
We use these boundary conditions because, in this particular case, 
 they completely eliminate the influence of boundaries 
on a numerical solution in internal points  (see below).
For integration in time we use
 MOL1(4)
with $cfl = \frac{1}{2}$. 
By varying  Courant number in the range $1 \leq cfl \leq \frac{1}{16}$ and 
Runge-Kutta 
order  from $n=2$ to $n=4$
we verified that errors of integration in time 
in our numerical experiments are
less than $10^{-4}$ of the
truncation errors  introduced by the spatial discretization \eqref{MOL1}. 

Results of numerical
 integration for a decaying solution $g_-$
are shown in Figure 2 for resolutions $N=16$ through $N=1024$.
In the beginning, numerical solutions 
follow the exact solution but eventually begin to deviate 
and grow exponentially.
With increasing $N$, the exponential growth starts later. However,
time $t_s$ of stable integration is proportional only to
a logarithm of a number of grid points, $t_s \sim \ln N$.
Integration beyond, say, $t\simeq 10$ would require 
an unpractical  larger number $N > 10^6$.
Table 3 illustrates convergence of numerical solutions for
two different moments of time, $t_1=1.2375$ and $t_2=3.7125$, which are 
both less than $t_s$. There is a second-order convergence at these times ( 
integration of $g_+$ 
does not present any difficulties and can be continued
indefinitely).


\bigskip
\begin{figure}
\centering
\includegraphics[totalheight=4in]{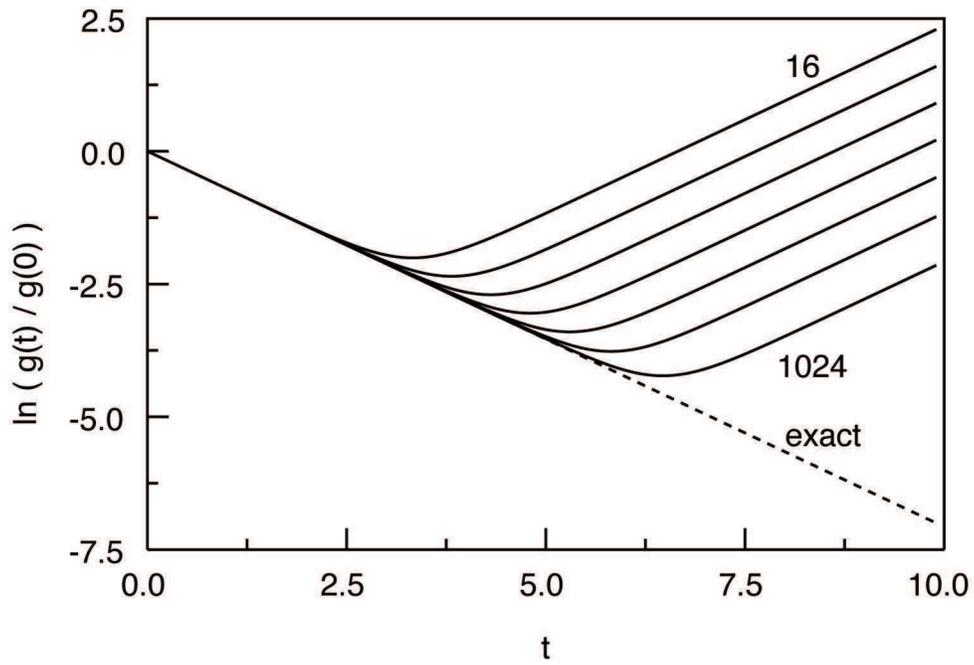}
\caption{A decaying solution $g_-$ as a function of time. 
Obtained using a MOL1(4) integrator
 and $cfl=1/2$. Solid lines - numerical solutions with $N=16$ through $N=1024$.
Dashed line - exact solution.}
\end{figure}


\begin{table}[!tbp]

\centerline{
\begin{tabular}{|r|r|r|}
\hline
N &  $L_{\infty}(t1)$ & $L_{\infty}(t2)$  \\
\hline
16 & 2.6421E-04 & 2.2272E-03    \\
32 &  5.6685E-05 & 4.4543E-04  \\
64 & 1.2938E-05 & 2.0090E-04  \\
128 & 3.0747E-06 & 5.9922E-05  \\
256 &7.3409E-07 & 1.5723E-05  \\
\hline
\end{tabular}
}

\caption{Convergence of numerical solutions $g_-$ for two moments of time
 $t_1=1.2375$ and $t_2=3.7125$.}

\bigskip

\end{table}

We now analyze the reason for an asymptotic instability 
 discussed above.
Using initial conditions $g_i(0) = exp(x_i)$, we can present 
numerical solutions $g_i(t)$ 
 at $t>0$ as
\labeq{SIMPLEQ1}{
                 g_i(t) = f_i(t) \exp ( x_i ),\quad 
                 K_i(t) = \pd{f_i}{t}\exp ( x_i ), 
}
where $f_i$ are functions of time.
 Substituting \eqref{SIMPLEQ1} into \eqref{MOL1}, we obtain a set of identical equations for
all $f_i$,
\labeq{SIMPLEQ2}{
\pdd{f_i}{t} = c f_i - \frac{(\pd{f_i}{t})^2}{f_i}
}
where 
\labeq{C}{
 c = \frac{e^{ \Delta x} + e^{- \Delta x} - 2}{\Delta x^2}
 = 1 + O(\Delta x^2) \geq 1~
}
is a constant independent of $i$ (Robin boundary conditions \eqref{ROBIN} are
necessary to ensure that \eqref{SIMPLEQ2} holds for 
the outermost grid points $i=1$ and $i=N$). 
Initial values $f_i(0) = 1$ and $\pd{f_i(0)}{t} = \pm \frac{1}{\sqrt{2}}$
are also identical
for all $i$. We thus can ignore index $i$ in \eqref{SIMPLEQ2}, and  use a
 single ordinary
differential equation 
\labeq{SIMPLEQ3}{
\pdd{f}{t} = c f - \frac{(\pd{f}{t})^2}{f}
}
to describe 
 numerical solutions $g_i(t) = f(t) \cdot g_i(0)$ at all grid points.
%
%
Integration  of  \eqref{SIMPLEQ3} gives
\labeq{SIMPLEQ5}{
    \left(\pd{f}{t}\right)^2 = \frac{c f^2}{2} + \frac{C}{f^2},
}    
where $C=const$. We set it to $C = \frac{1-c}{2}\leq 0$ to satisfy
 initial conditions, and finally obtain the following differential equation
\labeq{SIMPLEQ6}{
         \frac{df}{dt} = \pm \sqrt{\frac{c f^2}{2} + \frac{1-c}{2f^2}}.
}
The value of $c=1$ corresponds to a 
continuum limit of $\Delta x \rightarrow 0$. 
Difference in the behavior of 
 numerical and analytical solutions comes from the presence of
 the term $\frac{1-c}{2f^2}$. Note, that this term is $O(\Delta x^2)$. 
Its variations do not change the second-order accuracy of the algorithm.
The term is negative since $c> 1$.

Consider first an exponentially growing solution  $g_+$. For this solution,
$\pd{f(0)}{t}> 0$. Therefore, we must initially take 
a $+$ sign in \eqref{SIMPLEQ6}. 
The expression under the square root in \eqref{SIMPLEQ6}
is positive at $t=0$ and it will only increase with increasing $f$ because 
the term
$\frac{1-c}{2f^2} \rightarrow 0$  
 when
$f\rightarrow \infty$. The relative difference between numerical and 
analytical solutions will tend to zero
when $t\rightarrow\infty$ as well. 

For a decaying solution $g_-$,   
asymptotic behavior of exact and numerical
solutions is qualitatively different. 
For the exact solution, we have 
$K\rightarrow 0$ and 
$f\rightarrow 0$ when $t\rightarrow\infty$. But from \eqref{SIMPLEQ5}
 we observe that for $c$ other than zero
 $f$ cannot become zero because $\left(\pd{f}{t}\right)^2$ in
\eqref{SIMPLEQ5} has a minimum $\pd{f}{t}=0$ at a finite 
 $f=f_s= (2 ( c-1 )/ c)^{1/4} > 0$. 
For a decaying solution, $\pd{f(0)}{t} <0$ and we 
must initially take $-$ sign in \eqref{SIMPLEQ6}. 
When $f$ reaches the value of $f_s$ we have $df/dt=0$ but the second derivative remains  positive
\labeq{}{\frac{d^2f}{dt^2} = 2 c f + (c-1)/f^3 > 0.
}
As a result,  the numerical solution switches  at $f=f_s$ from $-$ branch to $+$ branch 
 in \eqref{SIMPLEQ6}, and  begins to increase, $f\rightarrow\infty$ 
when  $t\rightarrow \infty$. 

We can estimate the moment of time, $t_s$, when the solution 
switches from the $-$ to 
the $+$ branch from
\labeq{Ts}{
              e^{-\frac{t_s}{\sqrt{2}} } 
       \simeq f_s =(2 ( c-1 )/ c)^{1/4} \simeq  \Delta x^{1/2}.
}
When the resolution is increased, $f_s$ becomes smaller and is reached at later time. 
But  \eqref{Ts} shows that
$t_s \propto (\ln \Delta x)^{-1}$.
To increase a period of stable integration $t_s$ for a 
decaying
solution, we must decrease $\Delta x$ (increase $N$)
exponentially!

If we impose boundary conditions other than
 \eqref{ROBIN}, deviations from the
exact solution in internal points grow as described by \eqref{SIMPLEQ6} 
only until a signal from the boundary reaches them. After that, 
interaction with the boundary leads to a violent instability and a 
termination of numerical calculations. The reason for an instability 
observed for a decaying solution E2 (Figure 1) 
appears to be the same.
Truncation errors lead to an imperfect balance of linear and non-linear terms, and  
to a deviation of a numerical solution from the exact one.
Subsequent interaction with the boundaries amplifies these deviations, 
and the calculation eventually terminates. 

\section{Integration in logarithmic variables}

We now show that  
logarithmic variables $\phi=\ln g$, $\psi=\phi_t$, and $\theta=\psi_x$
allow to  significantly 
improve the accuracy and stability of numerical integration.

Equation \eqref{SIMPLEX} for $\phi$ can be integrated numerically
using the same schemes CFLN1 and MOL1$(n)$. We must simply 
 replace 
$g$, $K$ and $D$ in these schemes with $\phi$, $\psi$, and 
$\theta$, and to substitute the non-linear term 
$\cal R$ with its
logarithmic counterpart $\cal S$  \eqref{SIMPLEXS}. 
The  schemes than become
\labeq{CFLN1L}{ {\rm CFLN1:}\quad\quad \left\{
\begin{split}
\quad & \bar \psi_i = \psi^{n-\half}_i +
  \Delta t \left(\frac{\phi^n_{i+1} - 2 \phi^n_i + \phi^n_{i-1}}{\Delta x^2}
   + {\cal S}(\psi^{n-\half}_i,\theta_i^n)\right) \quad(\rm predictor), \\
\quad &  \psi_i^{n+\half} = \bar \psi_i + 
   \frac{ \Delta t}{2} \left( {\cal S}(\bar \psi_i,\theta^n_i)
                       - {\cal S}(\psi^{n-\half}_i,\theta^n_i)\right)
        \quad (\rm corrector) , \\
\quad &  \phi^{n+1}_i = \phi^n_i + \Delta t \, \psi^{n+\half}_i,\\
\end{split}
\right. 
}
and
\labeq{MOL1L}{ {\rm MOL1(n):} \quad\left\{
\begin{split}
\quad & \pd{\phi_i}{t} = \psi_i, \\
\quad & \pd{\psi_i}{t} = \frac{\phi_{i+1} + \phi_{i-1} - 2 \phi_i}{\Delta x^2} 
              + {\cal S}(\psi_i,\theta_i),
\end{split}
\right. 
}
where 
\labeq{}{
\theta_i = \frac{\phi_{i+1} - \phi_{i-1}}{2\Delta x}
}
 and
\labeq{}{
{\cal S}_i = - ( \alpha+1) \psi_i^2 - (\beta-1) \theta_i^2 
              - \gamma \psi_i\theta_i .
}
Schemes CFLN2 and MOL2 can be rewritten in a similar way.

Let us first consider how 
a CFLN1 scheme will reproduce an exponentially decaying
 solution $g_-$ \eqref{E3}.
For this solution, initial conditions are
$\phi^0_i = x_i$, a linear function of $x$, and
 $\psi^{-\half}_i = \frac{1}{\sqrt{2}}$,  a constant.
With these conditions, the right-hand sides 
of the first two equations in
 \eqref{CFLN1L} become zero, and  it is easy to verify
 that $\psi_i$ remain
 constant through all subsequent time steps, whereas $\phi_i$
decrease  with time linearly,  $\phi^n_i = x_i -\frac{ n \Delta t}{\sqrt{2}}$.
The same is obviously true for
 MOL1. Transformed to logarithmic variables, CFLN1 and MOL1
reproduce exponential solutions exactly.

\bigskip
\begin{figure}
\centering
\includegraphics[totalheight=4in]{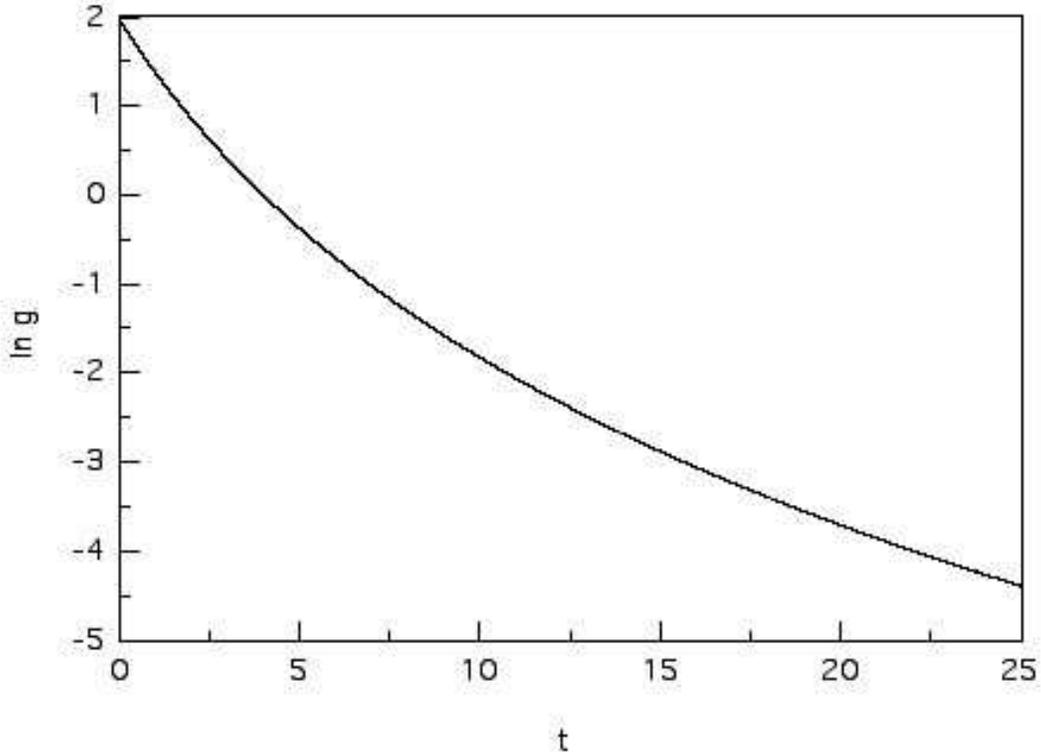}
\bigskip
\bigskip
\bigskip
\bigskip
\caption{Decaying solutions \eqref{E2} at $x=0.6$ as a function of time
obtained in logarithmic variables
using CFLN1 scheme with $cfl=1$. 
 Solid lines - numerical solutions for $N=128$ and  $N=2048$. 
Dashed line - exact solution. 
The solutions cannot be distinguished on the plot.
}
\end{figure}


\begin{table}[!tbp]

\centerline{
\begin{tabular}{|r|r|r|r|r|}
\hline
N &  $L_{\infty}(t1)$ CFLN1 & $L_{\infty}(t2)$ CFLN1 & $L_{\infty}(t1)$ MOL1 & $L_{\infty}(t2)$ MOL1\\
\hline
32   & 3.9E-02 & 8.9E-02 & 4.1E-02 & 2.1E-01 \\
64   & 1.3E-02 & 3.2E-02 & 1.3E-02 & 3.4E-02 \\
128  & 3.4E-03 & 1.1E-02 & 3.7E-03 & 1.3E-02 \\
256  & 8.7E-04 & 3.4E-03 & 9.1E-04 & 3.6E-03 \\
512  & 2.1E-04 & 8.9E-04 & 2.1E-04 & 9.0E-04 \\
1024 & 5.2E-05 & 2.2E-04 & 5.2E-05 & 2.2E-04 \\
\hline
\end{tabular}
}

\caption{Convergence of numerical solutions \eqref{E2}  
for two moments of time
 $t_1=20$ and $t_2=50$ using logarithmic 
variables and two numerical schemes, CFLN1 with  CFL number $cfl=1$ and 
MOL1(4) with $cfl=0.5$.  The error norm 
 $L_\infty = {\rm max}_i \vert g_i / g_e - 1 \vert$,
 where $g_e$ is the exact solution. }

\bigskip

\end{table}

Next, we test the reformulated  schemes on 
 a decaying solution
\eqref{E2}. Figure 3 shows numerical solutions at $x=0.6$ as a function of $t$ obtained using CFL1 scheme,
 and it must be compared to Figure 1. 
Table 4 illustrates convergence of reformulated
 CFLN1 and MOL1 schemes with increasing $N$ and should be compared 
with Table 2.
We see from the comparison that logarithmic variables greatly 
reduce errors of numerical integration.

\section{Conclusions}

In this paper we introduced a scalar wave equation with non-linear terms
 similar to those of more complex equations 
of general relativity. The equation has a number of non-trivial analytical 
solutions useful for testing numerical schemes.
 
We formulated two classes of
 finite-difference  schemes for numerical integration of
this equation. One (CFLN) is a non-linear extension of a 
classical
second-order central difference scheme for a linear scalar wave equation
 \cite{CFL1928}.
Another (MOL) is based on a method-of-lines approach. The schemes 
have a comparable
accuracy but MOL requires a larger number of right-hand side evaluations. 

Both schemes converge 
to exact solutions at any fixed $t$ when numerical resolution is increased, 
$\Delta x\rightarrow 0$. For some of the solutions, however, integration 
becomes
unstable when resolution is kept fixed and $t$ is increased.
We trace this behavior to deviations from a perfect balance between
linear  and non-linear terms caused by discretization. 
As a result, the asymptotic behavior of numerical solutions
 may differ qualitatively from the asymptotic 
behavior of the corresponding exact solutions of a partial differential 
equation. An important point is that these deviations 
happen without violating a second-order accuracy. Therefore, having 
both
convergence at finite $t$ and an asymptotic instability is not a 
contradiction.

An asymptotic instability seems not to be a fault of a particular
 numerical scheme.  
All schemes tested
in this paper display this phenomenon regardless 
of how numerical solutions are advanced in time. CFLN, Runge-Kutta, and 
ICN-type integration results in the same asymptotic instability. 
We  have no reason to believe that this phenomenon
should be limited only to scalar non-linear equations.
Examples presented in the paper clearly  demonstrate
 that second-order accuracy of spatial
discretization, although necessary for obtaining convergent numerical 
solutions at finite $t$, is not a guarantee of  a correct asymptotic behavior
of a numerical scheme.

Finally, we have shown that an
 exponential transformation \eqref{SUBEXP} leads to
a significant improvement in  accuracy and stability of numerical algorithms.
We do not know if some
 of the difficulties encountered in numerical general relativity
steam from  a similar non-linear
asymptotic instability, and whether integration of GR equations
can be improved by an exponential transformation of variables. 
We believe that these questions are worth investigating.
 In the accompanying paper we
show how an exponential transformation of variables 
can be  carried out for tensorial equations of GR. 
Numerical schemes formulated in this paper can be 
extended to GR equations written in both second-order (CFLN1 and MOL1)
and first-order forms (CFLN2 and MOL2).

\bigskip
\noindent
This work was supported in part by the NASA grant SPA-00-067, 
Danish Natural Science Research Council through grant No 94016535, 
Danmarks
Grundforskningsfond through its support for establishment of the Theoretical Astrophysics Center,
 and by the Naval Research Laboratory
through the Office of Naval Research. We thank  Kip
Thorne, Jacob Hansen, and Ryoji Takahashi for useful discussions. I.D. thanks the Naval Research Laboratory,
A.K. thanks the
 Theoretical Astrophysics Center,  and both authors thank Caltech for hospitality
during their visits.


\end{document}